\documentclass[%
 reprint,
 amsmath,amssymb,superscriptaddress,
 aps,
 prd,
 floatfix
]{revtex4-2}

\usepackage{graphicx}
\usepackage{float}
\usepackage{physics}
\usepackage[dvipsnames]{xcolor}
\usepackage[shortlabels]{enumitem}
\usepackage{bigints}

\usepackage[normalem]{ulem}

\newcommand{\calL}{{\cal L}}
\newcommand{\calO}{{\cal O}}

\newcommand{\hmu}{\, {\hat{\mu}}}

\newcommand{\lat}[2]{{{#1}^3\!\!\times\!\!{#2}}}

\allowdisplaybreaks
\begin{document}

\title{Finite density QCD phase structure from strangeness fluctuations}

\author{Szabolcs Bors\'anyi}
\affiliation{Department of Physics, Wuppertal University, Gaussstr.  20, D-42119, Wuppertal, Germany}

\author{Zolt\'an Fodor}
\affiliation{Pennsylvania State University, Department of Physics, State College, PA 16801, USA}
\affiliation{Pennsylvania State University, Institute for Computational and Data Sciences, University Park, PA 16802, USA}
\affiliation{Department of Physics, Wuppertal University, Gaussstr.  20, D-42119, Wuppertal, Germany}
\affiliation{Institute  for Theoretical Physics, ELTE E\"otv\"os Lor\' and University, P\'azm\'any P. s\'et\'any 1/A, H-1117 Budapest, Hungary}
\affiliation{J\"ulich Supercomputing Centre, Forschungszentrum J\"ulich, D-52425 J\"ulich, Germany}

\author{Jana N. Guenther}
\affiliation{Department of Physics, Wuppertal University, Gaussstr.  20, D-42119, Wuppertal, Germany}

\author{Piyush Kumar}
\affiliation{Department of Physics, Wuppertal University, Gaussstr.  20, D-42119, Wuppertal, Germany}

\author{Paolo Parotto}
\affiliation{Dipartimento di Fisica, Universit\`a di Torino and INFN Torino, Via P. Giuria 1, I-10125 Torino, Italy}

\author{Attila P\'asztor}
\affiliation{Institute  for Theoretical Physics, ELTE E\"otv\"os Lor\' and University, P\'azm\'any P. s\'et\'any 1/A, H-1117 Budapest, Hungary}
\affiliation{MTA-ELTE Lendület "Momentum" Strongly Interacting Matter Research Group, Budapest, Hungary}

\author{Chik Him Wong}
\affiliation{Department of Physics, Wuppertal University, Gaussstr.  20, D-42119, Wuppertal, Germany}

\date{\today}

\begin{abstract}

Charting the phase diagram of Quantum Chromodynamics (QCD) at large 
density is a challenging task due to the complex action problem in 
lattice simulations. 
Through simulations at imaginary baryon chemical potential $\mu_B$ we 
observe that, if the strangeness neutrality condition is imposed, both the 
strangeness chemical potential $\mu_S/\mu_B$ and the strangeness susceptibility 
$\chi_2^S$ take on constant values at the chiral transition for varying $\mu_B$.
We present new lattice data to extrapolate contours of constant $\mu_S/\mu_B$ or $\chi_2^S$ to finite 
baryon chemical potential.
We argue that they are good proxies for the QCD crossover 
because, as we show, they are only mildly influenced by criticality and by 
finite volume effects. We obtain continuum limits for these proxies up to 
$\mu_B = 400$~MeV, through a next-to-next-to-leading order (N$^2$LO) Taylor 
expansion based on large-statistics data on $\lat{16}{8}$, $\lat{20}{10}$ and 
$\lat{24}{12}$ lattices with our 4HEX improved staggered action. 
We show that these are in excellent agreement with existing results for the chiral 
transition and, strikingly, also with analogous contours obtained with the
hadron resonance gas (HRG) model. On the $\lat{16}{8}$ lattice, we carry out the 
expansion up to next-to-next-to-next-to-next-to-leading order~(N$^4$LO), and 
extend the extrapolation beyond $\mu_B=500$~MeV, again finding perfect agreement 
with the HRG model. This suggests that the crossover line constructed from this 
proxy starts deviating from the chemical freeze-out line near 
$\mu_B\approx500$~MeV, as expected but not yet observed.
\end{abstract}

\maketitle

\section{Introduction}

The phase structure of strongly interacting matter has been the subject
of intense research in decade-long experimental programs 
at the SPS and LHC (CERN), SIS18 (GSI) and RHIC (BNL) accelerator facilities. 
There is clear evidence that heavy-ion collision experiments reach high enough temperatures for the 
quark gluon plasma phase to be created~\cite{STAR:2024bpc}. In this large 
temperature phase quarks are deconfined and chiral symmetry is approximately restored (up to 
the small explicit breaking due to finite quark masses). 
The transition from hadronic to quark matter for zero quark-antiquark asymmetry 
(i.e., zero baryon chemical potential $\mu_B$) and physical quark masses is a 
smooth crossover, as was shown using finite-size scaling in lattice QCD 
simulations~\cite{Aoki:2006we}, located at around 
$T = 156-158$~MeV~\cite{Bazavov:2018mes,Borsanyi:2020fev}. 
At larger chemical potentials, a rich phase diagram is conjectured. In 
particular, the search for the critical endpoint of QCD has received
special attention, both by theory and experiment~\cite{An:2021wof}.

One way to locate the QCD crossover temperature on the lattice is by finding a 
peak in the susceptibility of the pseudo-order parameter, namely the chiral 
susceptibility~\cite{Borsanyi:2010bp,Bazavov:2011nk}, as a function of 
the temperature.
In the limit of vanishing quark masses (the so-called chiral limit), this 
susceptibility is associated to the second order chiral transition,
and follows the critical behavior of the three dimensional $O(4)$ universality 
class~\cite{Pisarski:1983ms}. The connection between QCD and the three-dimensional $O(4)$ spin model has been intensively studied on the 
lattice~\cite{Ejiri:2009ac,Ding:2019prx}, but also with Dyson-Schwinger 
equations (DSE)~\cite{Gao:2021vsf,Bernhardt:2023hpr} and the functional renormalization 
group (FRG)~\cite{Braun:2023qak}. On the lattice, the crossover temperature as a 
function of the baryon chemical potential has been calculated by means of Taylor 
expansion \cite{Bonati:2018nut,Bazavov:2018mes} and analytic continuation from purely imaginary 
$\mu_B$~\cite{Bonati:2015bha,Bellwied:2015rza,Bonati:2018nut, HotQCD:2018pds, Borsanyi:2020fev}, based on the 
location of the peak of the chiral susceptibility, providing results up to 
around $\mu_B = 300$~MeV~\cite{Bazavov:2018mes,Borsanyi:2020fev}. 
The theoretical community distinguishes between cross-over lines depending on the handling of the strange quark. Two frequently used options are the theoretically simpler choice
of vanishing strangeness chemical potential ($\mu_S=0$) and the strangeness neutrality ($n_s=0$), which is motivated by
the experimental setup of heavy ion collisions. The latter
case is more involved for theory, because $\mu_S$ has to be tuned to meet the neutrality condition.
The differences between the two setups have been often emphasized and quantified by the lattice community \cite{Bonati:2015bha,Ding:2024sux}.

An experimental proxy of the crossover temperature (valid at small-enough 
chemical potentials) is given by the temperature of chemical freeze-out, namely 
the stage of a heavy-ion collision where the chemical composition of the 
resulting hadronic medium is fixed (up to final-state decay processes). Since this 
can by definition only happen in the hadronic phase, the chemical freeze-out 
temperature naturally represents a lower bound on the QCD transition 
temperature. Still, due to the rapid drop in scattering rates at the crossover, 
it is expected to be very close to the actual crossover temperature, especially 
for large collision energies~\cite{Braun-Munzinger:2003htr}. Indeed, recent 
estimates at LHC energies show a freeze-out temperature that agrees with the 
crossover determined on the lattice within errors~\cite{Andronic:2018qqt}. 
As the chemical potential increases, the freeze-out curve approaches the nuclear 
liquid-gas transition~\cite{Floerchinger:2012xd}, and is thus no longer expected 
to follow the crossover line, but rather to deviate downward and separate from 
it. In the range of chemical potentials where lattice QCD calculations for the 
crossover line exist, this deviation has not yet been 
observed~\cite{Borsanyi:2025lim, HotQCD:2018pds, Borsanyi:2020fev}. 

Determining the crossover line up to large density and where it deviates from 
the freeze-out line is crucial for our understanding of the phase structure of 
QCD, as well as for the interpretation of experimental measurements, which 
provide snapshots of fluctuation observables taken at freeze-out conditions. 
The freeze-out curve can also serve as a lower 
bound on the critical point location~\cite{Lysenko:2024hqp}, at least if one 
assumes that the chiral critical endpoint is also a deconfinement critical 
endpoint.
Moreover, the critical endpoint should be located on the analytic continuation 
of the crossover line.

In order to extend the lattice QCD phase diagram to higher chemical potentials 
(so that a possible deviation from the freeze-out curve can be observed) it is useful to 
study other proxies for the crossover temperature, instead of the theoretically 
cleaner definition based on the chiral susceptibility. One unfortunate aspect 
of the chiral susceptibility is that it suffers from sizable finite-volume 
effects. On the one hand, this makes extrapolations to large chemical potentials 
very difficult, as signal-to-noise ratios in Taylor coefficients deteriorate 
exponentially with the volume. On the other hand, this volume dependence is 
unsurprising, since this observable effectively uses light quarks (with a large 
Compton wavelength) to probe the medium. Recently, in 
Ref.~\cite{Borsanyi:2025lim} finite-volume effects have been compared for 
observables based on light quarks and infinitely heavy (static) quarks, and 
indeed it was observed that the values of static quark observables have much smaller finite-volume corrections.
However, observables based on static quarks have different undesirable 
properties. First, compared to the chiral pseudo-order parameter (the chiral 
condensate) the Polyakov-loop (which is used to define the static quark free 
energy) is noisier. Second, the slope of the static quark free energy is small 
and weakly dependent on the temperature. As a consequence, the static quark 
entropy has an extremely broad peak. Indeed, it was estimated that, while at 
$\mu_B=0$ the width of the chiral transition is around 
$\sim 15$~MeV~\cite{Borsanyi:2020fev}, the width of the deconfinement transition 
is much larger, around $\sim 35$~MeV~\cite{Borsanyi:2024xrx}. In a way, the 
Polyakov loop and related observables are less sensitive to the QCD crossover.

Thus, light quark observables have a larger volume dependence, while infinitely 
heavy quark observables have a smaller volume dependence, but are also less 
sensitive to the crossover. It is then natural to ask what happens if one 
probes the medium at the scale of the intermediate mass quark: the strange quark. 

In this work we observe that two strangeness-related
quantities are remarkably constant along the chiral transition line, at imaginary
as well as real chemical potentials, as long as the strangeness neutrality condition is satisfied. 
The first is the strangeness susceptibility $\chi^S_2$, defined as the second 
derivative of the QCD pressure with respect to the strange quark chemical 
potential, or the grand canonical variance of strangeness:
\begin{equation}
\chi^S_2 = \frac{\partial^2 \left( p/T^4 \right)}{\partial (\mu_S/T)^2} \, \, ,
\end{equation}
where $p$ is the pressure, $T$ is the temperature and $\mu_S$ is the strangeness 
chemical potential. 
The second is the strangeness chemical potential needed to reach strangeness neutrality, normalized by the baryon chemical potential, namely $\mu_S/\mu_B$. 

The strangeness susceptibility has some technical aspects that are advantageous 
in lattice calculations. First, unlike static quark quantities, the slope of 
$\chi^S_2$ near $T_c$ is large enough to make the calculation of 
$\chi^S_2 \approx $~const. curves practical. Second, due to the larger strange 
quark mass, the associated remnant of the sign problem is milder, as compared to 
quantities dominated by light quarks, making it possible to extrapolate to 
larger chemical potentials. Third, like the static quark quantities, and unlike 
light-quark-based quantities, the strangeness susceptibility has a remarkably 
mild volume dependence (at least at $\mu_B=0$). The ratio $\mu_S/\mu_B$ is 
directly determined by the strangeness density, and enjoys the same advantages 
as the strangeness susceptibility. Thus, if we can successfully argue that the 
conditions $\chi^S_2, \mu_S/\mu_B = {\rm const.}$ are good proxies for the 
crossover, this will allow us to chart the phase diagram to an unprecedentedly 
large chemical. We note that the observation that $\mu_S/\mu_B \approx {\rm const.}$ 
along the crossover line is not new, as was already shown in 
Ref.~\cite{Bollweg:2024epj} for real chemical potentials from a Taylor expansion.

The observation upon which this work is based, that both the strangeness 
susceptibility $\chi^S_2$ and the ratio $\mu_S/\mu_B$ are remarkably constant at 
the crossover temperature as the chemical potential is increased, is an 
empirical statement based on lattice data at zero and purely imaginary chemical 
potentials. 
Using strange quarks, instead of light quarks to probe the hot and dense QCD 
medium allows us to use a smaller volume, and thus higher statistics. 

In this work we employ data from three different lattices, all with aspect ratio 
$LT=2$ and $N_\tau = 8,10,12$. By means of a two dimensional Taylor expansion in $\mu_B,\mu_S$ 
we obtain a controlled extrapolation of the different contours, and provide the 
continuum limit of our two proxies of the chiral transition up to 
$\mu_B \approx 400$~MeV. We observe mild cut-off dependence, especially in 
$\mu_S/\mu_B$. We also find finite volume effects to be small, by comparing finite
density results in two volumes, with $LT=2$ and $LT=3$.
Remarkably, we find that the 
corresponding results from the hadron resonance gas (HRG) model agree with our 
results for both observables and at all chemical potentials.
The extreme statistics we accumulated on our $\lat{16}{8}$ lattice
allows us to obtain, for this single lattice spacing, the proxies for the QCD 
transition up to $\mu_B \approx 550$~MeV. In this case, the 
$\mu_S/\mu_B = {\rm const.}$ curve detaches from the parametrized chemical 
freeze-out line from Ref.~\cite{Andronic:2017pug} above $\mu_B=400$~MeV, while 
remaining in perfect agreement with the HRG model prediction.

The structure of the paper is as follows. In the next section, we present 
phenomenological arguments for why the strangeness susceptibility may be a good 
proxy for the crossover line. This argumentation has two steps.
First, we show directly using lattice QCD data (at real and several different 
purely imaginary chemical potentials, as well as several volumes) that, at small 
chemical potentials, constant values of the strangeness susceptibility or of 
the ratio $\mu_S/\mu_B$ indeed coincide with the peak of the chiral 
susceptibility. We also show that, unlike the chiral susceptibility, $\chi^S_2$ 
has a very weak volume dependence, which is practically useful if one wants to 
accumulate the extreme statistics needed for a very high order Taylor expansion. 
Second, we show (using a phenomenological calculation) that the critical 
fluctuations of strangeness are suppressed when strangeness neutrality 
is imposed on the system. This means that such a correspondence should break 
down much more slowly when approaching the critical point than for other 
observables: e.g. no such suppression of critical fluctuations takes place for 
the chiral susceptibility. In Section~\ref{sec:lattice} we first discuss our 
lattice setup, then present our results for the finite $\mu_B$ extrapolations of 
the contours and finally their continuum limit. We compare our results with 
other relevant lines in the phase diagram, such as the chiral crossover and the 
chemical freeze-out line from heavy ion collisions phenomenology. Finally, we 
summarize our results and discuss their relevance in mapping the phase diagram 
of QCD in Section~\ref{sec:discussion}.

\section{Strangeness fluctuations near the phase boundary\label{sec:chisatc}}

We start by arguing that the lines with a constant strangeness susceptibility 
$\chi^S_2$, or with a constant $\mu_S/\mu_B$, are good proxies for the crossover 
line on the QCD phase diagram. We have two arguments: the first is empirical and 
based on lattice simulations at imaginary chemical potential; the second is 
phenomenological and applies in the vicinity of a supposed critical endpoint. 

\subsection{Strangeness fluctuations as proxies for the QCD crossover}

\begin{figure}
    \centering
    \includegraphics[width=\linewidth]{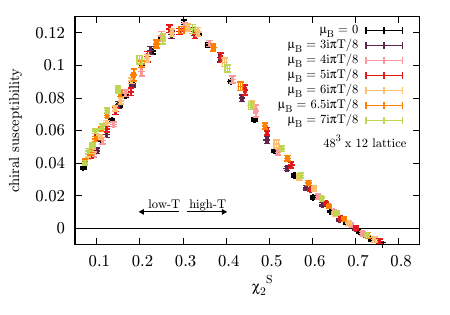}
    \includegraphics[width=\linewidth]{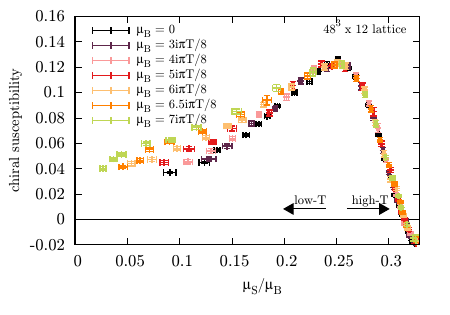}
    \caption{The chiral susceptibility in the case of strangeness neutrality, as a function of the strangeness susceptibility (top) and the normalized strangeness chemical potential (bottom), on a $\lat{48}{12}$ lattice for different imaginary values of the baryon chemical potential. }
    \label{fig:pbpsusc_vs_SS}
\end{figure}

Our initial observation is based on lattice QCD data sets perviously used in 
Refs.~\cite{Borsanyi:2020fev,Borsanyi:2022qlh,Borsanyi:2025dyp}. These 
simulations use lattices of size $48^3 \times 12$ with the 4stout improved 
staggered action~\cite{Bellwied:2015lba}. Here we do not present new simulation 
results, but rather use different visualizations of already published results to 
make a point.

\begin{figure}
    \centering
    \includegraphics[width=\linewidth]{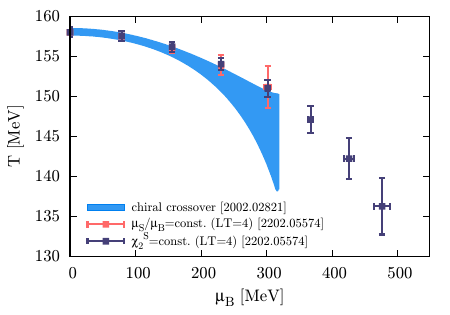}
    \caption{The contour obtained through the conditions $\chi^S_2={\rm const.}$ or $\mu_S/\mu_B={\rm const.}$ in continuum extrapolated results in the $T^\prime$ expansion from Ref.~\cite{Borsanyi:2022qlh}, compared to the chiral transition line from Ref.~\cite{Borsanyi:2020fev}.}
    \label{fig:4stout_Tc_vs_SScontour}
\end{figure}

We show the chiral susceptibility as a function of the strangeness 
susceptibility and of the ratio $\mu_S/\mu_B$, for several different imaginary 
chemical potentials in Fig.~\ref{fig:pbpsusc_vs_SS}. With the exception of the 
low-temperature regime for $\mu_S/\mu_B$, the different imaginary 
chemical potentials neatly fall on the same curve, and the position of the 
peaks is independent of $\mu_B$. The observed lattice data point to
apparently universal values for these strangeness-related variables, defined
by the peak of the chiral susceptibility:
$\chi^S_2 = {\rm const.} \approx 0.3$ and
$\mu_S/\mu_B = {\rm const.} \approx 0.25$.
Note that Fig.~\ref{fig:pbpsusc_vs_SS} shows a broad range of imaginary chemical potentials, 
$\mu_B/T=0 \dots i\cdot 2.75$. The largest chemical potential is close to $\mu_B/T=i\pi$,
where the Roberge-Weiss critical point is located~\cite{Roberge:1986mm}. The fact that we observe this apparent collapse even close to this point
indicates that the associated critical region is narrow, and non-singular behaviour dominates.
We do not have data exactly at $\mu_B/T=i\pi$ because of the numerical challenge posed
by setting the strangeness neutrality condition in the presence of critical slowing down.

The observation in the imaginary domain of the chemical potential implies that, 
after analytic continuation, the curves defined by the peak of the chiral susceptibility, and
those defined by proxies ($\chi^S_2 = {\rm const.} \approx 0.3$ and
$\mu_S/\mu_B = {\rm const.} \approx 0.25$) are bound to agree, at least at small 
enough real chemical potentials. This is seen in 
Fig.~\ref{fig:4stout_Tc_vs_SScontour}, where we compare the chiral transition 
curve calculated in Ref.~\cite{Borsanyi:2020fev} from analytic continuation 
using a polynomial ansatz and the curves of constant $\chi^S_2$ or $\mu_S/\mu_B$ 
obtained using the $T'$-expansion in Ref.~\cite{Borsanyi:2022qlh}. Up to around 
$300$~MeV, all bands agree within error.

\begin{figure}
    \centering
    \includegraphics[width=\linewidth]{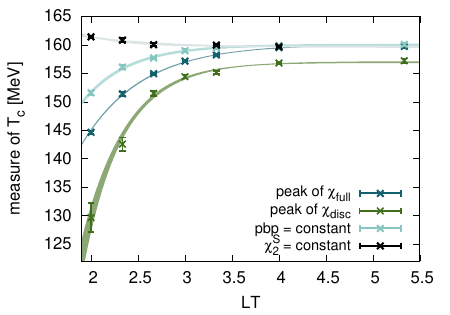}
    \caption{Different definitions and proxies of the QCD crossover temperature as functions of the lattice volume in temperature units (the aspect ratio) on $N_\tau=12$ lattices. 
    \label{fig:manytc_vol}}
\end{figure}

Collapse plots similar to Fig.~\ref{fig:pbpsusc_vs_SS} have been presented for 
the chiral susceptibility versus the chiral condensate in 
Ref.~\cite{Borsanyi:2020fev}. In fact, a constant value of the chiral condensate 
could also be suggested as a proxy of the crossover temperature. However, that 
choice has some technical/practical disadvantages. 
One important disadvantage is that definitions of $T_c$ based on the chiral 
condensate of chiral susceptibility have larger finite volume effects than those 
based on the strangeness susceptibility (at least at zero chemical potential)~\cite{Borsanyi:2025lim}. 
This is shown in Fig.~\ref{fig:manytc_vol}, where different proxies for $T_c$ 
are shown as functions of the aspect ratio $LT$ (the spatial extent of the 
lattice in temperature units). A smaller simulation volume helps to reach higher 
chemical potentials, as the signal-to-noise ratios of the Taylor coefficients 
strongly deteriorate with volume~\cite{Adam:2025hpb}. Furthermore, the per-configuration-cost is 
also smaller for smaller volumes, which makes it possible to gather 
the very large statistics needed for a high order Taylor expansion.

We remark that similar collapse plots also underlie the rationale for the $T'$ 
expansion, which is a resummation of the Taylor series designed to converge 
quickly~\cite{Borsanyi:2021sxv, Borsanyi:2022qlh, Abuali:2025tbd} in situations 
where such collapse is a good approximation. This happens if the free energy 
density is to a good approximation a single-variable function of the combination 
$T(1-\kappa \hmu_B^2)$\footnote{Here we employed the compact notation $\hmu_i=\mu_i/T$.}, with $\kappa$ a constant, instead of $T$ and $\mu_B$ 
separately. This appears to be a good assumption at small chemical potentials. 
However, it is also an assumption that should break down at higher chemical 
potentials if the critical endpoint exists.

\subsection{The (in)sensitivity of strangeness fluctuations to the QCD critical endpoint}

So far our arguments were based on lattice QCD data, and we only argued that a 
constant value of the strangeness susceptibility or $\mu_S/\mu_B$ are good 
proxies for the crossover temperature at small chemical potentials. 
In general, such contours starting from the pseudo-critical 
temperature at $\mu_B=0$ will hit the critical point if: 1) they follow the 
chiral transition line, and 2) the value of the observable they are based on is 
not influenced by critical effects. 

We already showed Fig.~\ref{fig:4stout_Tc_vs_SScontour} in support of argument 
1), and will present more precise, continuum extrapolated lattice results in a
smaller physical volume in the next Section. As far as argument 2) is concerned,
we note that the value of $\mu_S/\mu_B$ needed for strangeness neutrality 
is driven by the strangeness density, which does not diverge at the critical 
point. On the other hand, $\chi_2^S$ is a susceptibility, which then could 
diverge at the critical point. However, we will show in the following that such 
divergence is strongly suppressed in the case of strangeness neutrality.

In fact, it is possible for fluctuation observables to be insensitive to the 
critical endpoint, if they do not couple to the baryochemical potential. That 
this is the case for e.g. net-pion fluctuations was pointed out a long time ago 
in the literature~\cite{Stephanov:1999zu, Hatta:2003wn}. This happens because, 
while the $\pi^+ \pi^+$, $\pi^+ \pi^-$ and $\pi^- \pi^-$ correlators are all 
singular at the critical point, due to isospin symmetry, and the pions not 
coupling to a baryochemical potential, these singular parts cancel in net-pion fluctuations.

This argument can also be applied to fluctuations of strange quarks, as we 
are going to show in this Section.
We will use a quark-meson model to show how the coupling to the critical 
$\sigma$ mode affects the expected critical contribution to $\chi_2^S$, both for 
$\mu_S=0$  (larger effect) and for strangeness neutrality $n_S=0$ (where it will 
turn out to be a much smaller effect). 
This will allow us to exploit, in the $n_S=0$ case, the empirical argument of 
$\chi_2^S = {\rm const.}$ at the chiral transition to estimate the location of 
the QCD phase boundary up to large $\mu_B$.

\begin{figure}
    \centering
    \includegraphics[width=\linewidth]{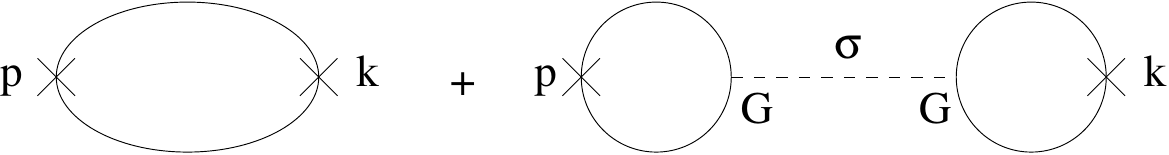}
    \caption{Diagrams contributing to the second order fluctuations of net-proton number. Figure from Ref.~\cite{Stephanov:1999zu}.}
    \label{fig:diagrams}
\end{figure}

Fluctuations of the net baryon number diverge at the critical point as powers of 
the system's correlation length~\cite{Stephanov:1999zu,Hatta:2003wn}: for example, the second baryon susceptibility diverges as:
\begin{equation}
    \chi_2^B \sim \xi^\frac{\gamma}{\nu} \, \, \rm,
\end{equation}
where $\xi$ is the correlation length, and $\gamma$ and $\nu$ are the critical
exponents for the 3D Ising universality class.

In an effective model treatment, this divergence is due to the coupling of baryons to the critical $\sigma$ field 
by means of a Yukawa term 
\begin{equation}
    \calL_{\sigma \bar{N} N} = G \sigma \bar{N} N  \, \, .
\end{equation}

This is conveniently reflected into experimentally available observables, such 
as event-by-event fluctuations and correlations of 
hadrons~\cite{Stephanov:1999zu}, in particular protons. Correlations between 
hadron species $i,j$ of order two can be written as:
\begin{equation} \label{eq:nncorr}
    \left\langle \Delta N^i \Delta N^j \right\rangle = V \bigintssss_p \bigintssss_k 
    \left\langle \delta n^i_p \delta n^j_k \right\rangle
\end{equation}
where $n_p^i$ is the occupation number of species $i$ in momentum mode $p$.

Contributions e.g. to the proton-proton correlator integrated over in 
Eq.~\eqref{eq:nncorr} are diagrammatically shown in Fig.~\ref{fig:diagrams}, 
where the critical contribution comes from the right diagram due to the exchange 
of a zero-momentum $\sigma$ field which becomes massless at the critical point.

The critical contribution to the second order net-proton fluctuations is then 
driven by the right diagram, which evaluates to~\cite{Hatta:2003wn}:
\begin{align} \nonumber
    V \left\langle \delta n_p \delta n_k \right\rangle &= \frac{g^2}{m_\sigma^2 T} \frac{4 m_p^2}{E_p E_k} \left[ n_p^+ \left( 1 - n_p^+ \right) - n_p^- \left( 1 - n_p^- \right) \right] \\
    & \qquad \quad \times \left[ n_k^+ \left( 1 - n_k^+ \right) - n_k^- \left( 1 - n_k^- \right) \right] \, \, ,
\end{align}
where $E_p = \sqrt{p^2 + m_p^2}$, $m_p$ is the proton mass, $m_\sigma$ is the 
diverging $\sigma$ mass, $T$ the temperature, 
$n_p^\pm = \left[ \exp{(E_p \mp \mu_p)/T} \right]^{-1}$ are the Fermi-Dirac 
distributions for proton and antiproton, and the factors $\left( 1 - n \right)$ 
account for the Pauli principle.

We now wish to estimate the size of the critical contribution to the strangeness 
susceptibility $\chi^{S {\rm(crit)}}_2$. In order to do
so, we apply the same line of reasoning to constituent quarks.
In this case, the Yukawa coupling term would be analogous
to the previous case:
\begin{equation}
    \calL_{\sigma \bar{q} q} = G \sigma \sum_{i=u,d,s} \bar{q}_i q_i  \, \, ,
\end{equation}
where we can assume the same coupling $G$ for all three quark flavours thanks to the $SU(3)_f$ symmetry.

This means that the macroscopic correlation between the net numbers of quarks of 
flavours $i,j$ reads:
\begin{equation} 
    \left\langle \Delta N^i \Delta N^j \right\rangle = V \bigintssss_p \bigintssss_k 
    \left\langle \delta n^i_p \delta n^j_k \right\rangle \, \, ,
\end{equation}
with, as before:
\begin{align} \label{eq:udscorr} \nonumber
    V \left\langle \delta n^i_p \delta n^j_k \right\rangle &= \frac{G^2}{m_\sigma^2 T} \frac{4 m_i m_j}{E^i_p E^i_k} \\
    \nonumber
    &\times \left[ n_p^{i ,+}\left( 1 - n_p^{i ,+} \right) - n_p^{i ,-} \left( 1 - n_p^{i ,-} \right) \right] \\
    & \times \left[ n_k^{j,+} \left( 1 - n_k^{j,+} \right) - n_k^{j,-} \left( 1 - n_k^{j,-} \right) \right] \, \, ,
\end{align}
where the $n^{i,\pm}_p$ are now the Dirac distributions for flavour $i$, and 
$m_i$ are the constituent quark masses $m_u = m_d = 340$~MeV or $m_s = 500$~MeV. 
This expression factorizes into:
\begin{equation} 
    V \left\langle \delta n^i_p \delta n^j_k \right\rangle = \frac{G^2}{m_\sigma^2 T} F^i_p F^j_k  \, \, ,
\end{equation}    
where we defined: 
\begin{equation}
F^i_p = \frac{2 m_i}{E^i_p} \left[ n_p^{i ,+}\left( 1 - n_p^{i ,+} \right) - n_p^{i ,-} \left( 1 - n_p^{i ,-} \right) \right]
\, \, .
\end{equation}

Apart from the explicit factor of temperature in Eq.~\eqref{eq:udscorr}, the 
whole dependence on $T,\mu_B,\mu_S$ is contained in $F^i_p$ through the quarks' 
Dirac distributions. The flavour-flavour correlation then reads:
\begin{align*}
     \left\langle \Delta N^i \Delta N^j \right\rangle = \frac{G^2}{m_\sigma^2 T} \bigintssss_p F^i_p \bigintssss_p F^j_p \, \, .
\end{align*}

In order to estimate the size of the critical contribution
to the strangeness susceptibility, consider the ratio:
\begin{equation} \label{eq:ratioR}
R (T,\mu_B) = \frac{\bigintssss_p F^u_p}{\bigintssss_p F^s_p} \, \, .
\end{equation}

By rearranging the derivatives with respect to the $u,d,s$ chemical potentials 
in terms of those related to the conserved charges $B,Q,S$, the relative size of 
the critical contributions to $\chi_2^S$ and $\chi_2^B$ becomes:
\begin{align} \label{eq:crit_ratio}
  \frac{\chi^{S {\rm(crit)}}_2}{\chi^{B {\rm(crit)}}_2} (T,\mu_B) &= \frac49 R (T,\mu_B)^2 + \frac49 R (T,\mu_B) + \frac19 \, \, .
\end{align}

We can estimate this ratio by considering values for $T_c,\mu_{B,c}$ consistent 
with current 
predictions~\cite{Fu:2019hdw,Gao:2020fbl,Gunkel:2021oya,Hippert:2023bel,Shah:2024img} 
for the critical point location, at $\mu_S=0$ or in the strangeness neutral 
condition, which as we showed roughly fixes the chemical potentials as 
$\mu_S \simeq 0.25 \mu_B$. For all such predictions, where $T=100-120$~MeV and 
$\mu_B=600-650$~MeV, from Eq.~\eqref{eq:crit_ratio} it follows that in the $\mu_S=0$ case the critical contributions satisfy 
$\chi_2^{S {\rm(crit)}} \simeq 0.3 \chi^{B {\rm(crit)}}_2$, while in the 
strangeness neutral case 
$\chi_2^{S {\rm(crit)}} \simeq 0.01 \chi^{B {\rm(crit)}}_2$.
This means that, in the latter case, the strangeness susceptibility $\chi^S_2$ 
is almost completely oblivious to the possible presence of a critical point. Because the constituent quark masses are generated by a finite expectation value of the $\sigma$ field, their value at the critical point will be lower than in the
vacuum. With the reasonable assumption that such expectation value at the critical point is reduced by a factor two~\cite{Schaefer:2007pw}, we can expect the light constituent quark masses to drop accordingly, and the strange quark mass to be reduced by the same (absolute) amount. In this simplified scenario, i.e. with  
constituent quark masses $m_u = m_d = 170$~MeV or $m_s = 330$~MeV, we obtain in the 
strangeness neutral case 
$\chi_2^{S {\rm(crit)}} \simeq 0.03 \chi^{B {\rm(crit)}}_2$.

Independently of the constituent quark masses, in the case of a zero \textit{strange quark} chemical potential $\mu_s=0$
one has $\chi_2^{S {\rm(crit)}} = 0$, as contributions from the strange 
quarks and their antiparticles cancel in this Feynman diagram exactly.

 \begin{figure*}[t]
     \centering
     \includegraphics[width=0.328\linewidth]{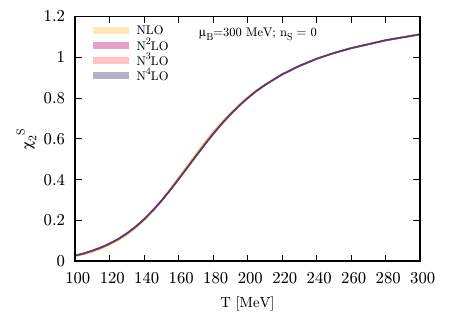}  
     \includegraphics[width=0.328\linewidth]{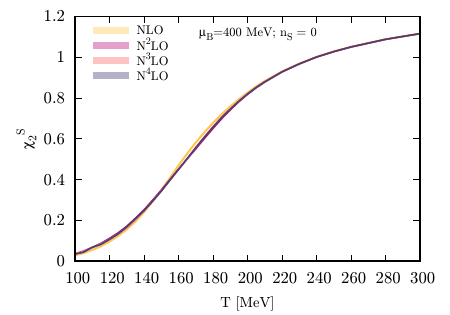} 
     \includegraphics[width=0.328\linewidth]{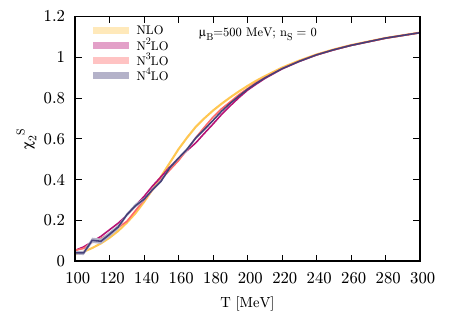}
     \caption{The extrapolated $\chi_2^S$ at different values of $\mu_B$ and different orders in the Taylor expansion, on our $\lat{16}{8}$ lattice. The strangeness chemical potential $\mu_S$ was set for each $T,\mu_B$ pair to match the
     strangeness neutrality condition. 
     \label{fig:SS_Taylor}
     }
\end{figure*}

\section{Lattice results at $LT=2$}
\label{sec:lattice}
We now briefly discuss the simulation setup for the new lattice results
we present on the $\chi_2^S$ and $\mu_S/\mu_B$ contours at finite chemical 
potential.

We use $N_f=2+1$ flavours of rooted staggered fermions with 4 steps of HEX 
smearing~\cite{Capitani:2006ni} and the DBW2 action \cite{QCD-TARO:1998nbk}, at 
physical values of the quark masses. We set the scale with the pion decay 
constant $f_\pi$, or with a modified version of the Wilson-flow-based 
$w_0$~\cite{BMW:2012hcm}, as introduced in Ref.~\cite{Borsanyi:2023wno}, where 
this particular lattice action was already used, as well as in 
Ref.~\cite{Borsanyi:2025lim}. 
We employ $\lat{16}{8}$, $\lat{20}{10}$ and $\lat{24}{12}$ lattices to perform 
continuum extrapolations of the transition line proxies up to $\mu_B=400$~MeV. 
We have here the same statistics as in Ref.~\cite{Borsanyi:2023wno} on the 
$\lat{20}{10}$ and $\lat{24}{12}$, while on the $\lat{16}{8}$ lattice 
it is much larger, the same as in Ref.~\cite{Borsanyi:2025lim}. 
Additionally, we employ new simulation results on a $\lat{24}{8}$ lattice to gauge finite 
volume effects (with 60000 - 70000 configurations per temperature). Thanks to the larger statistics, we can extrapolate our results 
on the $\lat{16}{8}$ lattice further in $\mu_B$, and provide guidance on where 
the transition might take place when the density is further increased.
On the smaller lattices ($\lat{16}{8}$ and $\lat{20}{10}$)
we use the reduced matrix formalism to calculate the fluctuations, in the same 
way as we did in Refs.~\cite{Borsanyi:2022soo, Borsanyi:2023tdp, Borsanyi:2023wno, Adam:2025phc},
on the larger lattices we determine the $\mu_B$-derivatives with stochastic sources \cite{Allton:2002zi}.

\subsection{Extrapolation to finite chemical potential}
\label{sec:Taylor}

In this work we take advantage of the extreme statistics we have gathered to 
build a two-dimensional Taylor expansion in $\mu_B,\mu_S$. We obtain results for 
the lines of constant $\chi_2^S$ or $\mu_S/\mu_B$ up to $\mu_B=400$~MeV, that we 
extrapolate to the continuum. This is based on an expansion up to 
next-to-next-to-leading (N$^2$LO) order in the chemical potentials. In our coarsest 
lattice $\lat{16}{8}$ we are able to employ coefficients up to 
next-to-next-to-next-to-next-to-leading (N$^4$LO) order, i.e. including up to tenth 
order conserved charge fluctuations and correlations, and push the extrapolation 
above $\mu_B = 500$~MeV. 

\begin{figure}
    \centering
    \includegraphics[width=\linewidth]{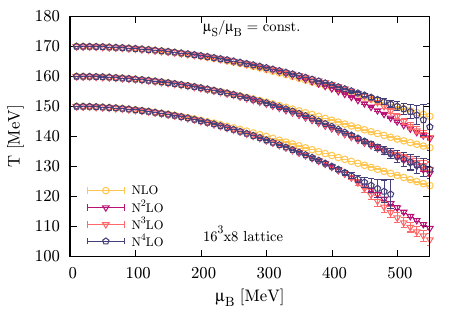}
    \includegraphics[width=\linewidth]{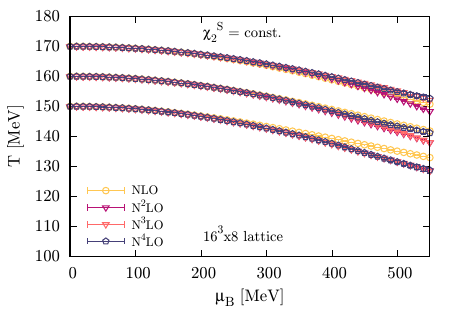}
    \caption{Contours of constant $\mu_S/\mu_B$ or $\chi_2^S$, on our $\lat{16}{8}$ lattice at different orders in the Taylor expansion.}
    \label{fig:contours_16x8}
\end{figure}

The results in this work refer to the strangeness neutral case, i.e. a setting of the 
strangeness chemical potential $\mu_S$ such that the strangeness density $n_S$ 
vanishes. We accomplish this by constructing the two dimensional Taylor expansion in 
$\mu_B,\mu_S$ and, for each $T$ and $\mu_B$, search for the 
$\mu_S = \mu_S^\star$ that corresponds to the strangeness neutral case.
This provides the value of $\mu_S^\star$ over the whole portion of the phase diagram 
we are able to access with our extrapolations. 
This scheme slightly differs from Ref.~\cite{Bazavov:2012vg} where $\mu_S^\star(\mu_B)$
itself is Taylor expanded. Our direct solution of -$n_S(\mu_S)=0$ at fixed $T$ and $\mu_B$
achieves a faster convergence in the orders of $\mu_B$. Once $\mu_S^\star(T,\mu_B)$ is known, we
can evaluate  $\chi_2^S (T,\mu_B,\mu_S^\star)$ at strangeness neutrality, too,
as shown in Fig.~\ref{fig:SS_Taylor}. Notice that, for all studied chemical
potentials, the strange susceptibility is a monotonic function of the
temperature. This means that the temperature where 
$\chi_2^S (T,\mu_B,\mu_S^\star) = \chi_2^S(T_c,0,0)$ is always well defined.

We show in Fig.~\ref{fig:contours_16x8} the extrapolated contours of constant 
$\mu_S/\mu_B$ (top) and $\chi_2^S$ (bottom) from our $\lat{16}{8}$ lattice at 
different orders in the Taylor expansion, from NLO to N$^4$LO, and observe very 
good convergence. The effects beyond N$^2$LO are visible only above 
$\mu_B \approx 450$~MeV, and a discrepancy between N$^3$LO and N$^4$LO appears, if 
at all, above $\mu_B \approx 500$~MeV.

\begin{figure}
    \centering
    \includegraphics[width=\linewidth]{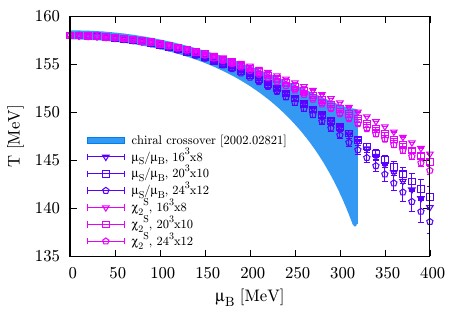}
    \caption{Contours of constant $\mu_S/\mu_B$ or $\chi_2^S$ from a N$^2$LO Taylor expansion, on our $\lat{16}{8}$, $\lat{20}{10}$ and $\lat{24}{12}$ lattices. Constant values are taken at $T_0=158$~MeV.}
    \label{fig:contours_Nt81012}
\end{figure}

\begin{figure}
    \centering
    \includegraphics[width=\linewidth]{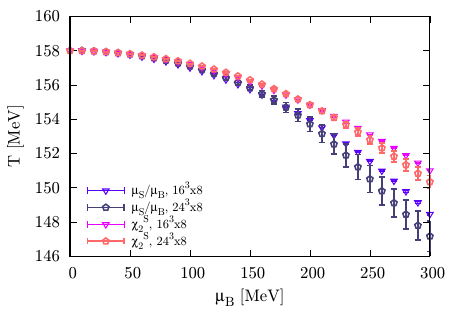}
    \caption{Contours of constant $\mu_S/\mu_B$ or $\chi_2^S$ from a N$^2$LO Taylor expansion, on our $\lat{16}{8}$ and $\lat{24}{8}$. Constant values are taken at $T_0=158$~MeV.}
    \label{fig:contours_comp_Nt8}
\end{figure}

In this work we are after proxies for the QCD transition line. In order to 
obtain a continuum limit, on each lattice we construct $\chi_2^S = {\rm const.}$ 
contours, where the constant is the $\mu_B=0$ value of $\chi_2^S$ at $T_0=158$~MeV, i.e. 
the result we obtained in the continuum in Ref.~\cite{Borsanyi:2020fev}. We then 
repeat the procedure for $\mu_S/\mu_B = {\rm const.}$ contours. The results are 
shown in Fig.~\ref{fig:contours_Nt81012}, where one can see that the two proxies 
yield slightly different results, although both are in agreement with the 
current continuum extrapolated result for the QCD 
crossover~\cite{Borsanyi:2020fev}, and their spread is in fact smaller that the error on 
such result. This result is based on a N$^2$LO Taylor expansion, i.e. including up 
to $\calO (\mu_B^6)$ contributions. We observe that discretization effects are 
smaller for the $\mu_S/\mu_B={\rm const.}$ contours, as no clear $N_\tau$ 
ordering appears, compared to the $\chi^S_2={\rm const.}$ contours.

Before carrying out the continuum limits, we wish to assess the size of finite 
volume effects. We perform the same analysis on a $\lat{24}{8}$ lattice, and 
compare to the $\lat{16}{8}$ result in Fig.~\ref{fig:contours_comp_Nt8}. We see 
that the results from both volumes are in good agreement in the whole range we 
can access with the statistics we gathered on our $\lat{24}{8}$ lattice. Hence,  
finite volume effects are smaller than discretization effects, and in particular 
smaller than the difference in temperature between the contours based on the 
different observables.  

\begin{figure*}
    \centering
    \includegraphics[width=\linewidth]{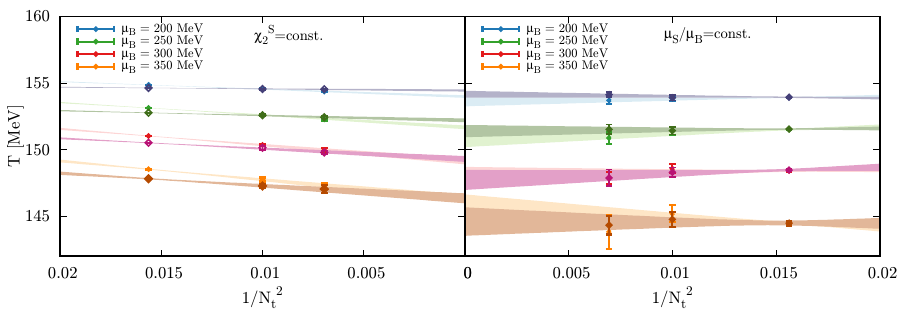}
    \caption{Continuum limits of the contours of constant $\chi_2^S$ (left) and 
    $\mu_S/\mu_B$ (right) at different values of $\mu_B$. Different points and bands in each plot indicate the two different scale settings we employ, with the lighter and darker bands corresponding to the $f_\pi$ and the $w_1$ scales, respectively.}
    \label{fig:contlim}
\end{figure*}

\subsection{Continuum limit}

\begin{figure}
    \centering
    \includegraphics[width=\linewidth]{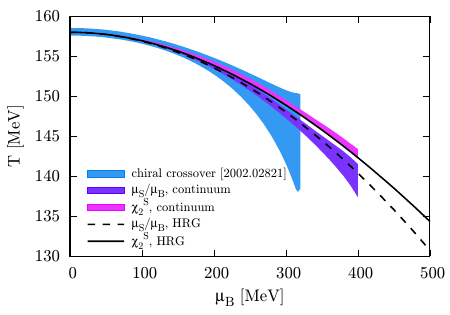}
    \caption{Contours of constant $\mu_S/\mu_B$ or $\chi_2^S$ from a N$^2$LO Taylor expansion (constant values taken at $T_0=158$~MeV), extrapolated to the continuum, compared to the chiral crossover and to HRG results.}
    \label{fig:cont_HRG}
\end{figure}

Finally, we perform the continuum extrapolation of the N$^2$LO Taylor expansion 
results up to $\mu_B=400$~MeV. In Fig.~\ref{fig:contlim} we show this for the 
strangeness susceptibility (left) and $\mu_S/\mu_B$ (right). We first observe, 
as noticed earlier, that cut-off effects are smaller for the latter, for which 
the results are almost $N_\tau$-independent. In both panels, the two sets of 
points and fit bands indicate the results obtained with the two scale settings 
we employ. These were defined and introduced for this action in Ref.~\cite{Borsanyi:2023wno}.
We consider the difference between the two scale settings as a source 
of systematic errors, and combine the two results in the following. We also 
observe that in general the continuum limits based on the two different 
observables are in good agreement, showing discrepancies around the $1\sigma$ 
level.

Combining the results from Fig.~\ref{fig:contlim} for each $\mu_B$ value (and 
including both scale settings), we obtain the continuum extrapolated proxies for 
the chiral crossover we show in Fig.~\ref{fig:cont_HRG}. Again, we find that 
these contours are in perfect agreement with the chiral crossover, and the 
tension between the contours based on the different observables is quite mild. 
The two contours could in fact be taken together, interpreting their spread as 
an additional source of systematic error on the QCD crossover. Even so, the 
error we obtain is much smaller than on the chiral crossover, thanks to the 
smaller physical volume and the extreme statistics we employed. Additionally, we 
show the corresponding contours obtained with the HRG model, and strikingly find 
that they are also in perfect agreement with our continuum extrapolations.
These are obtained in the same way as our lattice-based ones: we fix the value 
of the observables at $T_0 = 158$~MeV, then construct the contours at all 
$\mu_B$ values. The agreement we observe means that, although the HRG might 
(slightly) disagree with lattice results at $\mu_B=0$, the $\mu_B$ dependence
is captured correctly by the model. This is highly not trivial, given the simple
assumptions the model lies upon. From what we observe, one could take the HRG 
result itself as a proxy of the QCD crossover, in which case it would be
possible to extend its predictions to even larger chemical potential.

\begin{figure}
    \centering
    \includegraphics[width=\linewidth]{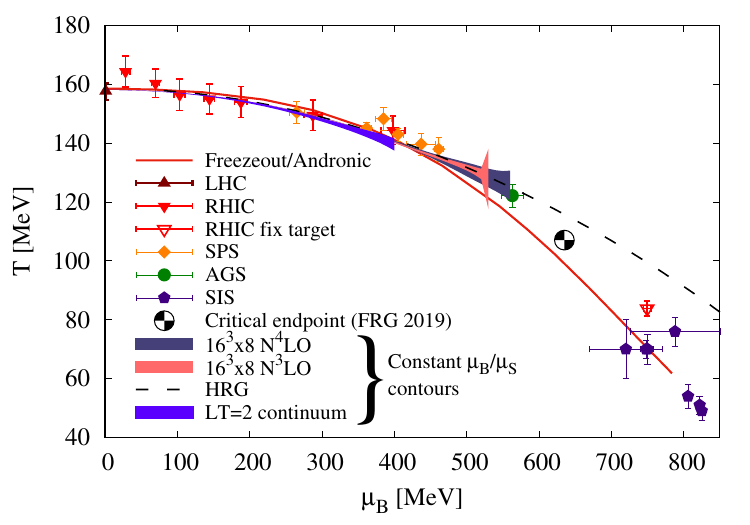}
    \caption{
    We show the contours of constant $\mu_S/\mu_B$ that satisfy
    the strangeness neutrality condition. On our coarsest lattice we give the
    $\mathcal{O}(\mu_B^6)$ (N$^3$LO) and $\mathcal{O}(\mu_B^8)$ (N$^4$LO)
    extrapolations of the contours up to 550 MeV, where the expansion seems to
    break down. The continuum limit (available at N$^2$LO) is computed up
    to 400~MeV. These are consistent with each other, but also with the
    prediction of the Hadron Resonance Gas model.
    We show in addition freeze-out data from various publications
    \cite{Vovchenko:2015idt,Becattini:2016xct,STAR:2017sal,Vovchenko:2018fmh,Lysenko:2024hqp}
    together with their parametrization of Ref.~\cite{Andronic:2017pug}
    and a recent functional result on the critical end-point's location \cite{Fu:2019hdw}.
    }
    \label{fig:muS_per_muB_result_plot}
\end{figure}

\section{Summary and discussion}
\label{sec:discussion}

In this work we discussed how observables related to strangeness fluctuations can 
shed light on the phase structure of QCD at finite density.
The starting point of our discussion was the remarkable collapse of the chiral susceptibility when
plotted against the strange susceptibility 
$\chi^S_2$ as the chemical potential was varied. This implied that
the chiral transition can be associated with a specific value of $\chi^S_2\approx 0.3$.
Similar statements are true for our other observable, the $\mu_S/\mu_B$ ratio,
satisfying the strangeness neutrality condition, though the data collapse is only observed at the high
temperature side of the transition. In the latter case $\mu_S/\mu_B\approx 0.25$ seems to characterize
the transition in a broad range of chemical potentials. Since direct simulations in large volumes are
only feasible at zero or imaginary chemical potentials, our initial observation was limited to this domain.

One may wonder at this point, why we can make this statement in the strangeness neutrality context only,
and if there are other thermodynamic variables that behave similarly. Obvious candidates would be the
normalized baryon density $n_B/\mu_B$ or the baryon susceptibility $\chi_2^B$. One of the
requirements for a successful proxy variable is that it is monotonic in
temperature. The monotonicity of $n_B/\mu_B$ and the corresponding data collapse have already been exploited
to construct the $T^\prime$-expansion in Refs.~\cite{Borsanyi:2021sxv,Borsanyi:2022qlh}. By the nature of
this construction, the $\chi^B_2$ is predicted to be non-monotonic, and this is confirmed by lattice data
at imaginary $\mu_B$. Its sensitivity to critical behaviour in the vicinity of either the
Roberge-Weiss end-point or the chiral end-point is obvious. Another criterion for a successful proxy,
though, is to have a $\mu_B$-independent Stefan-Boltzmann (SB) limit. While $n_B/\mu_B$ is a sigmoid in temperature for a broad range of chemical
potentials, its high temperature (SB) limit is $\mu_B$-dependent, and one cannot associate a universal
value with the ``middle of the transition''. The same is true for $\chi^S_2$, unless the strangeness neutrality condition is imposed (or, as an approximation, the fixed ratio $\mu_S=\mu_B/3$ is considered, which corresponds to a vanishing \textit{strange quark} chemical potential 
$\mu_s=0$).
We argued that the same condition suppresses the coupling of this proxy to the critical fluctuations near the CEP.
Considering also the practical advantages of defining the proxy on the scale of the strange mass, we are
left with the two proxies, $\chi^S_2$ and $\mu_S/\mu_B$, both constrained with the phenomenologically relevant
strangeness neutrality condition.

Next, we extended the initial observation to real chemical potentials, by comparing to the chiral transition line we have already published in 
Ref.~\cite{Borsanyi:2020fev}. To extrapolate the proxies themselves, we evaluated them via the $T^\prime$-expansion with the continuum extrapolated coefficients of Ref.~\cite{Borsanyi:2022qlh}. We found good
agreement with the chiral line, though with large errors. One may argue that this comparison is trivial,
since both the proxies and the chiral observables were in agreement at imaginary $\mu_B$, and however sophisticated 
their analytical continuation may be, they are bound to remain equal.
For this reason here we followed a different strategy. First, we argued that the
finite volume corrections on our chosen observables are under control even if we
reduce the simulation volume to two inverse temperatures ($LT=2$). 
The practical advantage of this is the availability of high order baryon
fluctuations, since the severity of both the sign and overlap problems is  exponential in the volume.
Armed with these generalized susceptibilities of baryon and strangeness we could calculate sufficiently high
orders of the Taylor expansion, and demonstrate that subsequent orders
are negligible up to a given $\mu_B$.
This range is $\mu_B\approx 400$~MeV if the highest available coefficient is N$^2$LO, but stretches out 
to $550$~MeV if we can afford the $N^4$LO coefficients. This latter range is what we could cover
with our extreme statistic ensembles at the coarsest lattice, while the former range applies to the continuum limit (driven down by the much higher cost of the $\lat{24}{12}$ lattices).
Even with this limitation, we could continuum extrapolate both of our proxies to a broader range of chemical
potentials than what is known today as the chiral crossover line.

It is not obvious that the coarsest lattice in the study is close to the continuum limit. With our discretization, we do observe this for the ratio 
$\mu_S/\mu_B$.
To highlight the implication of our results for current knowledge of the QCD phase diagram, we show a sketch of the latter covering a broad range in temperature and chemical potential in Fig.~\ref{fig:muS_per_muB_result_plot}.
The continuum extrapolated
contour of constant $\mu_S/\mu_B$, starting at $T_c$ at $\mu_B=0$, is shown together with the same contour for two subsequent
orders on the $\lat{16}{8}$ lattice. The highest order corresponds to employing up to 10$^{\rm th}$ order fluctuations.
$\chi^B_{10}$ was first presented in our recent work \cite{Adam:2025phc}, and the analogous strange derivatives
are used in this work for the first time. This unprecedented high order extrapolation allows us to confidently
predict strangeness-related observables at finite density at least as far as this lattice size allows.

Quite remarkably, this observable is in agreement with the hadron resonance gas model's prediction in the
entire range where the contour was computed (also shown in Fig.~\ref{fig:muS_per_muB_result_plot}).
Equally remarkable is the fact that, assuming the validity of the proxies introduced here, the HRG model can
predict the crossover.  We stress that the
only input we used for the HRG-based contour is the starting temperature at $\mu_B=0$, that we get from lattice QCD to be $T_0=158$~MeV. 

Given the experimental knowledge of the chemical freeze-out (see data points in Fig.~\ref{fig:muS_per_muB_result_plot})
one can estimate the point of divergence between the freeze-out line and the QCD crossover. 
We stress that
we have not demonstrated the validity of our proxies for the entire $\mu_B$ range. However, if we assume that they work, and that finite volume and discretization effects on 
$\lat{16}{8}$ lattice data are under control, we may be guided by the HRG prediction and predict that the two curves start deviating between 400 and 500~MeV. 

A similar insight may come from the freeze-out data themselves.
Along the parametrization of the freeze-out line from Ref.~\cite{Andronic:2017pug}, which we show in Fig.~\ref{fig:muS_per_muB_result_plot} as a red curve, the $\mu_S/\mu_B$
ratio maintains a near-constant value up to $\mu_B\approx400$~MeV,
before dropping to lower values at higher densities.
This drop in the value of $\mu_S/\mu_B$ might indicate
the point of deviation from the cross-over line. On the other hand, it might 
also signal the breakdown of this proxy. In any case, we see no sign of 
breakdown up to 400~MeV, which is the range where a continuum result is 
provided in this work.

Finally, the location of the critical end-point, for which first principles 
theoretical predictions outside of lattice already
exist~\cite{Fu:2019hdw,Gao:2020fbl,Gunkel:2021oya}, is expected to be in the range $\mu_B \sim 600 - 650$~MeV.
We argued that these proxies are weakly influenced by critical behaviour, implying that the
critical point is expected to be close to the lines defined by the constant values of $\chi^S_2$ or
of $\mu_S/\mu_B$. While the two proxies may diverge at high $\mu_B$, the range spread by
these two contours is still the best estimate for the transition line that
lattice QCD can offer today.

\begin{acknowledgments}
S. B. and P. P. thank Jan M. Pawlowski for fruitful discussions on the subject of the paper.
This work is supported by the
MKW NRW under the funding code NW21-024-A.
Z. Fodor acknowledges funding from the DOE under the contract number DE-SC0025025.
This work was also supported by the
Hungarian National Research, Development and Innovation
Office, NKFIH Grant No. KKP126769.
This work was also supported by the NKFIH excellence
grant TKP2021{\textunderscore}NKTA{\textunderscore}64.
This work is also supported by the Hungarian National Research,
Development and Innovation
Office under Project No. FK 147164.
The authors gratefully acknowledge the Gauss Centre for
Supercomputing e.V. (\url{www.gauss-centre.eu}) for funding
this project by providing computing time on the GCS Supercomputer
Juwels-Booster at Juelich Supercomputer Centre.
We acknowledge the EuroHPC Joint Undertaking for awarding this project access to the EuroHPC supercomputer LUMI, hosted by CSC (Finland) and the LUMI consortium through a EuroHPC Extreme Access call.
An award of computer time was provided by the INCITE program. 
This research used resources of the Argonne Leadership Computing Facility, which
is a DOE Office of Science User Facility supported under Contract DE-AC02-06CH11357.
\end{acknowledgments}

\bibliographystyle{unsrt}
\bibliography{thermo,paolo}

\end{document}